\documentclass[preprint,3p,times,showcase]{elsarticle}

\usepackage[utf8]{inputenc}
\usepackage[T1]{fontenc}
\usepackage{amsmath,amssymb,color,graphicx,bm}
\usepackage{times}
\usepackage{dcolumn}
\usepackage{marginnote,colordvi}

\newcommand{\bal}{\boldsymbol{\alpha}}
\newcommand{\bbe}{\boldsymbol{\beta}}
\newcommand{\bga}{\boldsymbol{\gamma}}
\newcommand{\bde}{\boldsymbol{\delta}}

\begin{document} 

\begin{frontmatter}

\title{(2+1)-dimensional KdV, fifth-order KdV, and Gardner equations derived from the ideal fluid model. Soliton, cnoidal and superposition solutions }

\author{Anna Karczewska \thanks{Corresponding author, e-mail: A.Karczewska@wmie.uz.zgora.pl} }
\address{Institute of Mathematics, University of Zielona G\'ora, Szafrana 4a, 65-516 Zielona G\'ora, Poland}
\author{Piotr Rozmej \thanks{ Professor Emeritus}}
\address{Institute of Physics, 
University of Zielona G\'ora, Szafrana 4a, 65-516 Zielona G\'ora, Poland}


\begin{abstract}
We study the problem of gravity surface waves for an ideal fluid model in the (2+1)-dimensional case. We apply a systematic procedure to derive the Boussinesq equations for a given relation between the orders of four expansion parameters, the amplitude parameter $\alpha$, the long-wavelength parameter $\beta$, the transverse wavelength parameter $\gamma$, and the bottom variation parameter $\delta$. We derived the only possible (2+1)-dimensional extensions of the Korteweg-de Vries equation, the fifth-order KdV equation and the Gardner equation in three special cases of the relationship between these parameters. All these equations are non-local. 
When the bottom is flat, the (2+1)-dimensional KdV equation can be transformed to the Kadomtsev-Petviashvili equation in a fixed reference frame and next to the classical KP equation in a moving frame. We have found soliton, cnoidal, and superposition solutions (essentially one-dimensional) to the (2+1)-dimensional Korteweg-de Vries equation and the Kadomtsev-Petviashvili equation.
\end{abstract}

\begin{keyword}
 Boussinesq's equations \sep  (2+1)-dimensional KdV fifth-order KdV and Gardner equations\sep Kadomtsev-Petviashvili equation\sep soliton, cnoidal and superposition solutions 
  \PACS 02.30.Jr\sep 05.45.-a\sep 47.35.B\sep 47.35.Fg
\end{keyword}

\end{frontmatter}


\section{Introduction}\label{intro}
The Korteweg-de Vries equation \cite{KdV} is one of the most widely used nonlinear wave equations. It was derived by perturbation theory from a model of an ideal fluid (non-viscous and incompressible) in which the motion is irrotational in (1+1)-dimension and the bottom of the fluid reservoir is flat. The simplest nonlinear wave equation in (2+1)-dimensions is the Kadomtsev-Petviashvili equation (KP for short) \cite{KP}. Both the KdV and KP equations are integrable and satisfy many conservation laws. Their analytical solutions are also known.

Since nonlinear waves on the surface of seas and oceans are very important in practice, many papers on (2+1)-dimensional equations of KdV or KP type can be found in the literature, see e.g. 
\cite{Wazwaz08,Wazwaz09,Peng10,Wazwaz11,WZZQ14,Adem16,Zhai16,Zhang17,Batwa18,Wang19,Lou20,Wazwaz20,Ozkan21,Cao21,Kumar22}. The equations used in these works  are also integral, which allows their authors to construct analytic solutions of various types (solitons, periodic solutions, lumps, breathers).

Extending the ideal fluid model to the (2+1)-dimensional case leads to much more complicated equations. 
The only such attempt known to us is erroneous, as we have shown in the work \cite{RK21Nody} (to avoid citation of the incorrect paper we do not cite it directly).
To the best of our knowledge, there are no correct papers that consider in full detail the (2+1)-dimensional waves in ideal fluid model. This is because the theory leads to equations that are non-integrable.

In this paper, we generalise the perturbation approach described in \cite{BS2013} and \cite{KRcnsns} for the (1+1)-dimensional case. The authors of \cite{BS2013} have shown that for a flat bottom, one can derive the KdV, extended KdV, fifth-order KdV, and Gardner equations for different relations between small parameters determining nonlinearity $(\alpha)$ and dispersion $(\beta)$. In \cite{KRcnsns}, we have generalized this approach for the case with an uneven bottom, assuming that the bottom changes are much smaller than the fluid's depth. Here, following methods used in \cite{KRcnsns} and \cite{KRnody22} we attempt to derive analogous wave equations for (2+1)-dimensional case. 

In \cite{KRnody22}, we have discussed four cases of dependence  between small parameters in detail. In all these cases, it was impossible to reduce the obtained Boussinesq equations to a single KdV-type equation for the wave profile. It was only possible to obtain a single wave-type nonlinear partial differential equation in which the argument is an auxiliary function determining the velocity potential. The solution of such an equation, if known, determines the (2+1)-dimensional function representing the time-dependent surface profile. In all these cases, however, the (2+1)-dimensional nonlinear wave equations are very complicated.

In this paper, we discuss three cases where the parameter $\gamma$ is in a higher order than the leading one, $\beta$ or $\alpha$. It turns out that for the flat bottom when $\alpha\approx\beta$ and $\gamma\approx\beta^{2}$ we can derive a true (2+1)-dimensional analogue of the KdV equation without any additional assumptions.
In two other cases, $\alpha\approx\gamma\approx\beta^{2}$ and $\beta\approx\gamma\approx\alpha^{2}$ we are able to derive (2+1)-dimensional analogues to fifth-order KdV and Gardner equations, respectively. 
 
The paper is organized as follows. In section \ref{descr},
to keep the article self-contained, we briefly repeat the description of the model  from \cite{KRnody22}. In section \ref{stand-g2} we derived 
the only true (2+1)-dimensional KdV equation. 
Next, in section \ref{simansol}, we showed the families of soliton, cnoidal and superposition solutions to this equation. 
Furthermore, if the bottom is equal, this equation trasformed to a moving reference frame implies the Kadomtsev-Petviashvili (KP) equation. In section \ref{5thEq}, for the case $\alpha\approx\gamma\approx\beta^{2}$ and flat bottom, we derived the Boussinesq equations. Next we showed that these Boussinesq's equations can be made compatible and reduced to a single (2+1)-dimensional fifth-order KdV equation. In section \ref{GardEq}, we considered the case $\beta\approx\gamma\approx\alpha^{2}$ and derived the (2+1)-dimensional Gardner equation. In section \ref{bottom}, we extended the (2+1)-dimensional fifth-order KdV and Gardner equations to the case with uneven bottom. Section \ref{concl} contains the conclusions.


\section{Description of the model} \label{descr}
Let us consider the inviscid and incompressible fluid model whose motion is irrotational in a container with an impenetrable bottom. 
In dimensional variables, the set of hydrodynamical equations consists of the Laplace equation for the velocity potential $\phi(x,y,z,t)$ and boundary conditions at the surface and the bottom.
\begin{eqnarray}  \label{g1}
\phi_{xx} + \phi_{yy} + \phi_{zz}&=& 0, \quad \mbox{in~the~volume}, \\ \label{g2}
\phi_z - ( u_x \phi_x +  u_y \phi_y + u_t) &=& 0, \quad \mbox{at~the~surface},  \\ \label{g3}
\phi_t + \frac{1}{2}(\phi_x^2+\phi_y^2+\phi_z^2) +g u &=& 0, \quad \mbox{at~the~surface}, \\ \label{g4}
\phi_z -h_{x}\phi_{x} -h_{y}\phi_{y} &=& 0, \quad \mbox{at~the~bottom}.
\end{eqnarray}
Here $ u(x,y,t)$ denotes the surface profile function, $g$ is the gravitational acceleration, and $\varrho$ is fluid's density. 
The bottom can be non-flat and is described by the function $h(x,y)$. 
Indexes denote partial derivatives, i.e.\ $\phi_{x}\equiv \frac{\partial \phi}{\partial x}, ~\phi_{yy}\equiv \frac{\partial^2 \phi}{\partial y^2}$, and so on. 

The next step consists in introducing a standard scaling to dimensionless variables (in general, it could be different in $x$-, $y$- and $z$-direction)
\begin{align} \label{bezw}
\tilde{x} & =x/L, \quad \tilde{y}= y/l,\quad \tilde{z}= z/H, \quad \tilde{t}= t/(L/\sqrt{gH}), \quad 
\tilde{ u}  =  u/A,\quad \tilde{\phi}= \phi /(L\frac{A}{H}\sqrt{gH}).\end{align}
Here, $A$ is the amplitude of surface distortions from equilibrium shape (flat surface), $H$ is average fluid depth, $L$ is the average wavelength (in $x$-direction), and $l$ is a wavelength in $y$-direction. In general,  $l$ should be the same order as $L$, but not necessarily equal. 
Then the set (\ref{g1})-(\ref{g4}) takes in scaled variables the following form (here and next, we omit the tilde signs)
\begin{align}  \label{G1}
\beta \phi_{xx} + \gamma\phi_{yy} + \phi_{zz}&= 0,  \\ \label{G2}
  u_t +\alpha ( u_x\phi_x+\frac{\gamma}{\beta} u_y\phi_y)-\frac{1}{\beta}\phi_z &= 0, 
 \quad \mbox{for~} z  =1+\alpha u, \\ \label{G3}
\phi_t + \frac{1}{2}\alpha \left(\phi_x^2+\frac{\gamma}{\beta}\phi_y^2+\frac{1}{\beta}\phi_z^2\right) +  u &= 0,  
 \quad \mbox{for~} z =1+\alpha u,  \\ \label{G4}
\phi_z - \beta\delta h_x\phi_x -\gamma\delta h_y\phi_y  &= 0, 
 \quad \mbox{for~} z =\delta h. 
\end{align}
Besides standard small parameters $\alpha=\frac{a}{H}$,  $\beta=\left(\frac{H}{L}\right)^2$ and $\gamma=\left(\frac{H}{l}\right)^2$, which are sufficient for the flat bottom  case, we introduced another one defined as $\delta=\frac{a_h}{H}$, where $a_h$ denotes the amplitude of bottom variations \cite{KRcnsns,KRI}.
In the perturbation approach, all these parameters, $\alpha,\beta,\gamma,\delta$, are assumed to be small but not necessarily of the same order.  
The standard perturbation approach to the system of Euler's equations (\ref{G1})-(\ref{G4}) consists of the following steps. First, the velocity potential is sought in the form of power series in the vertical coordinate
\begin{equation} \label{Szer}
\phi(x,y,z,t)=\sum_{m=0}^\infty z^m\, \phi^{(m)} (x,y,t),
\end{equation}
where ~$\phi^{(m)} (x,y,t)$ are yet unknown functions. The Laplace equation (\ref{G1}) determines $\phi$ in the form which involves only two unknown functions with the lowest $m$-indexes, $f(x,y,t):=\phi^{(0)} (x,y,t)$ and $F(x,y,t):=\phi^{(1)} (x,y,t)$ and their space derivatives. Hence,
\begin{align} \label{Szer1}
 \phi(x,y,z,t) & =\sum_{m=0}^\infty \frac{(-1)^m}{(2m)!} z^{2m}\, (\beta\partial_{xx}+\gamma\partial_{yy})^m  f(x,y,t)  
 + \sum_{m=0}^\infty \frac{(-1)^m}{(2m+1)!} z^{2m+1}\,  (\beta\partial_{xx}+\gamma\partial_{yy})^{m}  F(x,y,t). 
\end{align}
The explicit form of this velocity potential reads as
\begin{align} \label{pot8}
\phi = f & -\frac{1}{2} z^2 (\beta f_{2x}+\gamma f_{2y}) 
+ \frac{1}{24} z^4 (\beta^2 f_{4x}+2\beta\gamma f_{2x2y}+\gamma^2f_{4y}) 
+ \cdots   \\ & 
+  z F-\frac{1}{6} z^3 (\beta F_{2x}+\gamma F_{2y}) 
+ \frac{1}{120} z^5 (\beta^{2} F_{4x}+2\beta\gamma F_{2x2y}+\gamma^{2} F_{4y})+ \cdots  \nonumber
\end{align}

Next, the boundary condition at the bottom (\ref{G4}) is utilized. For the flat bottom case, it implies $F=0$, simplifying substantially next steps. In particular, $F=0$  makes it possible to derive the Boussinesq equations up to arbitrary order. 
For an uneven bottom, the equation (\ref{G4}) determines a differential equation relating $F$ to $f$. This differential equation can be resolved to obtain $F(f,f_{x},f_{xx},h,h_{x})$ but this solution can be obtained only up to some particular order in leading small parameter. Then, the velocity potential is substituted into kinematic and dynamic boundary conditions at the unknown surface (\ref{G2})-(\ref{G3}). Retaining only terms up to a given order, one obtains the Boussinesq system of two equations for unknown functions $ u,f$ valid only up to a given order in small parameters. The resulting equations, however, depend substantially on the ordering of small parameters.

In 2013, Burde and Sergyeyev \cite{BS2013} demonstrated that for the case of (1+1)-dimensional and the flat bottom, the KdV, the extended KdV, fifth-order KdV, and Gardner equations can be derived from the same set of Euler's equations (\ref{G1})-(\ref{G4}). Different final equations result from the different ordering of small parameters and consistent perturbation approach up to first or second order in small parameters.

 In 2020, we extended their results to cases with an uneven bottom in \cite{KRcnsns}, but still in (1+1)-dimensional theory.  We showed that the terms originating from the bottom have the same universal form for all these four nonlinear equations. However, the validity of the obtained generalized wave equations is limited to the cases when the bottom functions are piecewise linear. On the other hand, the corresponding sets of the Boussinesq equations are valid for the arbitrary form of the bottom functions. 

In the present paper there are four small parameters, $\alpha,\beta,\gamma,\delta$. In order to make calculations easier, we will follow the idea from \cite{BS2013,KRcnsns}, relating all small parameters to a single one, called \emph{leading  parametr}. This method allows easier control of the order of different terms, but the final forms of the resulting equations are presented in original parameters $\alpha,\beta,\gamma,\delta$. 
In \cite{KRnody22} we discussed the first four cases, listed in the Table \ref{tab1}. The table does  not contain all possible second-order cases, but only those that lead to well-known KdV-type and Gardner equations when reduced to (1+1)-dimensions.

\begin{table}[hbt] \caption{Some cases of different ordering of small parameters.} \label{tab1} \begin{center} 
\begin{tabular}{||c|c|c|c|c|c||} \hline Case & \hspace{1ex}$\alpha$\hspace{1ex} & \hspace{1ex}$\beta\hspace{1ex}$ & \hspace{1ex}$\gamma$ \hspace{1ex} &\hspace{1ex}$\delta$\hspace{1ex} & (1+1), $\delta=0$ \\  \hline 
1 & $O(\beta)$ & lead   & $O(\beta)$ & $O(\beta)$  &  KdV\\ \hline 
2 & $O(\beta)$ &  lead  & $O(\beta)$& $O(\beta^2)$  & ext KdV\\ \hline 
3 & $O(\beta^2)$ & lead   &$O(\beta^{2})$ & $O(\beta^2)$  &  fifth-ord KdV\\ \hline 
4 & lead  & $O(\alpha^2)$  & $O(\alpha^{2})$ & $O(\alpha^{2})$  & Gardner \\ \hline 
5 & $O(\beta)$  & lead  & $O(\beta^{2})$ & $O(\beta)$  & KdV \\ \hline 
6 & $O(\beta^{2})$  & lead  & $O(\beta^{2})$ & $0$  & fifth-ord KdV\\ \hline 
7 & lead  & $O(\alpha^2)$  & $O(\alpha^{3})$ & $0$  & Gardner \\ \hline 
\end{tabular} \end{center} 
\end{table}  
In Table \ref{tab1} the abbreviations are used, {\it lead} means leading parameter, {\it (1+1)} means (1+1)-dimensions, {\it ext KdV} means the extended KdV.

Cases 1-4 were recently discussed by us in \cite{KRnody22} where Boussinesq's equations were derived for all of them. 
In these cases $\gamma$ parameter was of the same order as the $\beta$ parameter, which caused that in the Boussinesq equation resulting from (\ref{G2}) the term  $\frac{\gamma}{\beta}f_{yy}$ appeared as zeroth-order one. This makes reduction of Boussinesq's equations originating from equations (\ref{G2})-(\ref{G3}) to a single KdV-type equation for the wave profile impossible. 
On the other hand, they can be reduced to a single, highly nonlinear partial differential equation for an auxiliary function $f(x,y,t)$, which determines the velocity potential but is not directly observed quantity. If known, the solution $f$ of this equation determines the surface elevation function. In practice, seeking solutions by this method seems to be extremely difficult \cite{KRnody22}.
Below, we study in detail cases 5-7. We will show that in these cases, because the term $\frac{\gamma}{\beta}f_{yy}$ does not appear in zeroth-order,
 it is possible to go further and derive (2+1)-dimensional nonlocal analogue to the KdV equation in first order perturbation approach.
In the following, we will show that in the second order perturbation approach, we can derive (2+1)-dimensional nonlocal analogues to fifth-order KdV equation and Gardner equation. 
It is worth emphasizing that this reduction of the Boussinesq equations to a single wave equation 
is possible only when $\gamma$ is of higher order than the leading parameter ($\beta$ or $\alpha$) because only in such case zeroth order Boussinesq's equations do not contain $y$-derivatives of the wave profile function (see Remark in section \ref{stand-g2}).

\section{The only possible extension of the KdV equation to (2+1)-dimensions \\
Case~5: \hspace{1ex}  $\bal=O(\bbe)$, \hspace{1ex} $\bga=O(\bbe^2)$, \hspace{1ex} $\bde=O(\bbe)$ 
} \label{stand-g2}

Heuristic derivations of the Kadomtsev-Pietviashvili  (KP) equation in \cite{KP} suggest that variations of the surface waves should be slower in the $y$-direction perpendicular to the wave propagation ($x$-direction). 
Therefore it is worth studying in detail the case when $~\gamma=O(\beta^2)$.
Denote now
\begin{equation} \label{gam2}
\alpha= A\,\beta, \quad \gamma = G\,\beta^2, \quad \delta= D\,\beta,
\end{equation}
where, as earlier, $A,G,D$ are arbitrary constants close to 1.

In this case, the boundary condition at the bottom (\ref{G4}) imposes the following relation
\begin{align} \label{r9c2}
F & = \beta^{2} D (h f_x)_x + \beta^{3}\left[DG (h f_y)_y +D^2  \frac{1}{2}\left(h^2 F_x\right)_x \right]  +O(\beta^{4}). \end{align}
Therefore, we can eliminate $F$ taking $~F=\beta^{2} D (h f_x)_x $ which is valid only up to second order in small parameters. So, the explicit form of velocity potential, valid up to second order, is now
\begin{align} \label{pot8c2}
\phi & = f -\frac{1}{2} z^2 (\beta f_{2x}+\gamma f_{2y}) 
+  z \beta^{2} D (h f_x)_x  + \frac{1}{24} z^4 \beta^2 f_{4x} .
\end{align}
Now, if we substitute the velocity potential (\ref{pot8c2}) into (\ref{G2}) and  (\ref{G3}), the resulting equations are valid only up to first order. 
Inserting the velocity potential (\ref{pot8c2}) into the kinematic boundary condition at the surface (\ref{G2}) and neglecting terms higher than the first order in small parameters yields
\begin{equation} \label{g2G2}
u_{t}+f_{xx} + \alpha (u f_{x})_{x} -\frac{1}{6} \beta f_{4x} +\frac{\gamma}{\beta} f_{yy} -\delta  (h f_{x})_{x} =0.
\end{equation}
Analogous steps with the dynamic  boundary condition at the surface (\ref{G3}) lead to the first order equation 
\begin{equation} \label{g2G3}
u+f_{t} + \frac{1}{2}\alpha f_{x}^{2} -\frac{1}{2} \beta f_{xxt}  =0.
\end{equation}
Equations (\ref{g2G2})-(\ref{g2G3}) constitute the first oder Boussinesq's equations for the case when $\alpha \approx \beta$, $\gamma \approx \beta^{2}$ and $\delta \approx \beta$, that is, for non-flat bottom. Despite the assumption that $\gamma$ is of the second order, the term $\frac{\gamma}{\beta} f_{yy} $ appears in the Boussinesq equation (\ref{g2G2}) as the first order one.  


Let us try to apply a standard method for making the Boussinesq equations (\ref{g2G2})-(\ref{g2G3}) compatible, which in (1+1)-dimensions leads to the Korteweg-de Vries equation.

By differentiating over $x$ the equation (\ref{g2G3}) and denoting $f_{x}\!=\!w, \hspace{1ex} f\!=\! \int\! w\, dx, \hspace{1ex} f_{yy}= \int\! w_{yy}\, dx $ we can write the equations (\ref{g2G2})-(\ref{g2G3}) in the form
\begin{align} \label{et}
u_{t}+w_{x} & + \alpha (u w)_{x} -\frac{1}{6} \beta w_{3x} +\frac{\gamma}{\beta}\! \int\! w_{yy}\, dx -\delta (h w)_{x} = 0,\\ \label{wt}
w_{t}+u_{x} & + \alpha w w_{x} -\frac{1}{2} \beta w_{xxt} =0.
\end{align}
Equation (\ref{et}) has a nonlocal form. When the problem is reduced to (1+1)-dimensions ($u,w$ not dependent on $y$)  
equations (\ref{et})-(\ref{wt}) reduce to the classical Boussinesq equations leading to the KdV equation. Note, that additional term $\frac{\gamma}{\beta} \int \! w_{yy}\, dx$ is the first order one because $\frac{\gamma}{\beta}\approx \beta$. Then in zeroth-order the following holds
\begin{equation} \label{0g2}
u_{t}+w_{x}=0, \quad w_{t}+u_{x} \quad \Longrightarrow \quad w=u, \quad w_{t}=-w_{x}, \quad u_{t}=-u_{x}.
\end{equation}
We seek such form of $w$ function which allows us to make equations (\ref{g2G2})-(\ref{g2G3}) compatible, that is reduce in first order to the same equation. We postulate $w$ in the following form
\begin{equation} \label{1w}
w =u+w^{(1)} = u +\alpha Q^{(a)}+\beta Q^{(b)}+\frac{\gamma}{\beta} Q^{(g)} +\delta Q^{(d)},
\end{equation}
where $\alpha Q^{(a)}, \beta Q^{(a)}, \frac{\gamma}{\beta} Q^{(g)}, \delta Q^{(d)}$ are first-order corrections. 
In principle, $w$ in (\ref{1w}) can contain one more first order correction, namely $\frac{\gamma}{\alpha}Q^{(ga)}$. The calculation shows, however, that $Q^{(ga)}=0$.

Now, we insert  $w$ given by (\ref{1w}) into (\ref{et})-(\ref{wt}) and retain terms up to first order. The result from  (\ref{et}) is 
\begin{equation} \label{et1}
u_{x} + u_{t} +\alpha \left(Q^{(a)}_{x} +2 u u_{x}\right) +\beta \left(Q^{(b)}_{x} -\frac{1}{6} u_{xxx}\right) + \frac{\gamma}{\beta}\left(Q^{(g)}_{x} \!+\! \int\! u_{yy}\, dx\,\right)+\delta\left(Q^{(d)}_{x} -(h u)_{x} \right) =0 .
\end{equation}
Fom (\ref{wt}) we obtain 
\begin{equation} \label{wt1}
u_{x} + u_{t} +\alpha \left(Q^{(a)}_{t} +u u_{x} \right) +\beta \left(Q^{(b)}_{t} -\frac{1}{2} u_{xxt}\right) + \frac{\gamma}{\beta}\left(Q^{(g)}_{t} \right) + \delta \,Q^{(d)}_{t} = 0.
\end{equation} 
Subtracting (\ref{wt1}) from (\ref{et1}) and replacing $t$-derivatives by $(-)x$-derivatives (thanks to (\ref{0g2}), 
$Q^{(a)}_{t} \to -Q^{(a)}_{x}, Q^{(g)}_{t} \to -Q^{(g)}_{x}, ~u_{xxt}\to - u_{xxx}$) we receive 
\begin{equation} \label{1cor}
\alpha (2 Q^{(a)}_{x} + u u_{x}) +\beta \left(2 Q^{(b)}_{x} -\frac{2}{3} u_{xxx}\right)+ \frac{\gamma}{\beta} \left(2 Q^{(g)}_{x}+\int u_{yy}\, dx\,\right)+\delta \left(Q^{(d)}_{x}-Q^{(d)}_{t} -(h u)_{x} \right) =0 .
\end{equation} 
Due to arbitrariness of small parameters (\ref{1cor}) is equivalent to four equations
\begin{equation} \label{pop1}
Q^{(a)}_{x} = -\frac{1}{2}u u_{x} , \qquad Q^{(b)}_{x} = \frac{1}{3} u_{xxx}, \qquad Q^{(g)}_{x} = - \frac{1}{2}\int u_{yy}\, dx ,
\end{equation}
and
\begin{equation} \label{pop1d}
Q^{(d)}_{x}-Q^{(d)}_{t} = (h u)_{x} .
\end{equation}
Integration of equations (\ref{pop1}) over $x$ yields 
\begin{equation} \label{pop1c}
Q^{(a)} = -\frac{1}{4}u^{2}, \qquad Q^{(b)} = \frac{1}{3} u_{xx}
\qquad Q^{(g)} = - \frac{1}{2} \int \left( \int u_{yy}\, dx\right) dx.
\end{equation}
These three first order corrections are common with the case of an even bottom.
Correction function $Q^{(d)}$ related to an uneven bottom allows for making the Boussinesq equations (\ref{et})-(\ref{wt}) compatible only when $h_{xx}=0$, that is, when the dependence of  bottom function $h(x,y)$ on $x$ is piecewise linear. 
In such a case one obtains
\begin{equation} \label{pop1dA}
Q^{(d)} = \frac{1}{4}\left(2 h u +h_{x}\int u\,dx\right).
\end{equation}
The detailed discussion of this problem in (1+1)-dimensions, for all four cases presented in Table~\ref{tab1} is contained in the article \cite{KRcnsns}.

Now, let us check that $w$ given by (\ref{1w}),(\ref{pop1c}),(\ref{pop1dA}), that is, 
\begin{equation} \label{pp1}
 w = u +\alpha\left( - \frac{1}{4} u^{2}\right) +\beta \left(\frac{1}{3} u_{xx}\right) +\frac{\gamma}{\beta}\left( - \frac{1}{2}\int (\int u_{yy}\, dx )dx\right)+\delta\left( \frac{1}{4}(2 h u +h_{x}\int u\,dx)\right), \end{equation}
 does indeed reduce the Boussinseq equations to the same wave equation (when restricted up to first order terms). After substituting $w$ into (\ref{et}) and leaving the terms up to first order, we obtain 
\begin{equation} \label{et11}
 u_{t} +u_{x} +\frac{3}{2} \alpha u u_{x}+ \frac{1}{6} \beta  u_{xxx} + \frac{1}{2} \frac{\gamma}{\beta} \int u_{yy}\, dx 
 - \frac{1}{4} \delta (2h u_{x}+ h_{x} u)=0.
 \end{equation}
The same action with  (\ref{wt}) gives (remember that $h_{xx}=0$)
\begin{equation} \label{wt11}
 u_{t} +u_{x} + \alpha (u u_{x} -\frac{1}{2}u u_{t})- \frac{1}{6} \beta  u_{xxt} - \frac{1}{2} \frac{\gamma}{\beta}\int \!\left( \!\int\! u_{yyt}\, dx\right) dx  + \frac{1}{4} \delta \left(2h u_{t}+ h_{x} \!\int\! u_{t} dx\right)
=0.
 \end{equation}
Since in zeroth-order $u_{t}=- u_{x}$, $u_{xxt}=-u_{xxx}$, $u_{yyt}=-u_{xyy}$  and $ \int\! u_{yyt}\, dx=-u_{yy} $, then  equation  (\ref{wt11}) does receive the same form as equation (\ref{et11}).

Equation (\ref{et11}) takes correctly into account an uneven bottom, but only under assumption that $h_{xx}=0$. In the case when the bottom is even, $\delta=0$, and the last term in (\ref{et11}) vanishes. In such case 
\begin{equation} \label{kdv21}
 u_{t} +u_{x} +\frac{3}{2} \alpha u u_{x}+ \frac{1}{6} \beta  u_{xxx} + \frac{1}{2} \frac{\gamma}{\beta} \int u_{yy}\, dx =0.
 \end{equation}
 
{\bf Equations (\ref{et11}) and (\ref{kdv21}) are indeed true  (2+1)-dimensional nonlocal analogues to the Korteweg-de Vries equation. They are derived from Euler's equations for ideal fluid model, when fluid's motion is irrotational, within first order perturbative approach.}\\[-2mm]

{\bf Remark} \textit{In \cite{KRnody22}, we considered cases when parameter $\gamma$ was the same order as $\beta$. In other words wavelength in $x$ and $y$ directions are of the same order. Then the term $\frac{\gamma}{\beta} f_{yy}$ appears already in zeroth order in the Boussinesq equation originating from kinematic boundary condition at the surface (\ref{G2}). Its presence causes that in zeroth order there are no relations (\ref{0g2}).  
The lack of such relations in zeroth order Boussinesq's equations makes the derivation of the single wave equation of (2+1)-dimensional KdV-type impossible}.

\section{Simple analytic solutions to (2+1)-dimensional KdV equation} \label{simansol}

In this section, we show that the (2+1)-dimensional KdV equation (\ref{kdv21}) has families of soliton, periodic cnoidal and superposition solutions. We follow the method applied by us in papers \cite{IKRR,RKI,RK}, assuming a hypothetical form of the solutions and finding conditions on the constants determining the solutions. It appears, however, that these solutions are essentially one-dimensional. The same types of solutions appear for the Kadomtsev-Petviashvili equation.


\subsection{Soliton solutions}

By direct calculation one can check that the function
\begin{equation} \label{ssol}
u=A\, \text{Sech}^{2} (k x+l y-\omega t) 
\end{equation}
 fulfils both (2+1)-dimensional KdV equation (\ref{kdv21}) and KP equation (\ref{kpt}) for arbitrary $k,l$  when 
\begin{equation} \label{A-omega}
A=\frac{4 k^{2} \beta}{3 \alpha}, \qquad \omega = k+\frac{2 k^{3} \beta}{3}+\frac{l^{2}\gamma}{2 k \beta}.
\end{equation}

\subsection{Periodic cnoidal solutions} \label{PCsol}
Let us check whether the equation (\ref{kdv21}) also possesses cnoidal solutions.

\subsubsection{Mathematical solution}
 Assume solutions in the form 
\begin{equation} \label{cn2}
u(x,y,t)=A\, \text{cn}^{2} [(k x+l y-\omega t),m].
\end{equation}
Inserting (\ref{cn2}) into (\ref{kdv21}), after simplifications, yields 
\begin{equation} \label{cn2a}
\frac{A\, \text{cn}\, \text{dn}\, \text{sn}}{3 \beta  k}
   \left[4 \beta ^2 k^4-8 \beta ^2 k^4
   m-6 \beta  k^2+6 \beta  k \omega -3 \gamma 
   l^2 +\left(12 \beta ^2 k^4 m-9 \alpha 
   A \beta  k^2\right)\text{cn}^2 \right]=0,
\end{equation}
where the arguments of the Jacobi elliptic function are omitted. Equation (\ref{cn2a}) imposes two conditions on coefficients $k,l,\omega,m,A$, namely
\begin{align}  \label{con1}
4 \beta ^2 k^4-8\beta ^2 k^4 m-6\beta k^2+6 \beta  k \omega -3\gamma l^2 &=0,
\\ \label{con2} \mbox{and} \hspace{25ex}
12 \beta ^2 k^4 m-9 \alpha  A \beta  k^2 & =0.
\end{align}
Then the cnoidal wave (\ref{cn2}) is the solution to (\ref{kdv21}) when
\begin{equation} \label{A-omega1}
A=\frac{4 k^{2} \beta}{3 \alpha} m, \quad \omega = k+\frac{2 k^{3} \beta}{3}(2m-1)+\frac{l^{2}\gamma}{2 k \beta} \quad \mbox{with~~arbitrary}\quad k,l \mbox{~~and~~} m\in (0,1] .
\end{equation}
Function $u(x,y,t)$ given by (\ref{cn2}) with the coefficients (\ref{A-omega1}) can be named a {\bf mathematical solution} to (2+1)-dimensional KdV equation (\ref{kdv21}).

\subsubsection{Physical solution}

The physical solution has to fulfill volume conservation. Therefore the physical solution has to have the form  
\begin{equation} \label{cn2B}
u(x,y,t)=A\, \text{cn}^{2} [(k x+l y-\omega t),m]+B,
\end{equation} 
where the value of $B$ has to ensure that the volumes of elevated and depressed fluid cancel over the intervals equal to wavelengths. This condition reads as
\begin{equation} \label{vcc} 
\int_{0}^{L_{x}} \int_{0}^{L_{y}}( A\, \text{cn}^{2} [(k x+l y-\omega t),m]+B)\, dx\,dy=0,
\end{equation}
where $L_{x},L_{y}$ are wavelengths in $x,y$ directions, respectively.
Periodicity of the $\text{cn}^{2}[(k x+l y-\omega t),m]$ function implies 
$$ L_x=\frac{2 K(m)}{k} \qquad \mbox{and} \qquad  L_y=\frac{2 K(m)}{l} ,$$
where $K(m)$ is the complete elliptic integral of the first kind.
So, (\ref{vcc}) gives
\begin{align} \label{vcc1}
B & =- \frac{A}{L_{x}\,L_{y}} \int_{0}^{L_{x}}\! \int_{0}^{L_{y}}  \text{cn}^{2} [(k x+l y-\omega t),m] 
= -\frac{A}{m }\left(\frac{E(m)}{K(m)}+m-1\right),
\end{align} 
where $E(m)$ is the complete elliptic integral.

When the form (\ref{cn2B}) is inserted into the (2+1)-dimensional KdV equation (\ref{kdv21}) the conditions imposed on coefficients (\ref{cn2a})-(\ref{con1})
become slightly modified. Instead (\ref{cn2a}) we obtain
\begin{align} \label{cn2b}
\frac{A\, \text{cn}\, \text{dn}\, \text{sn}}{3 \beta  k} 
   \bigg[   &  4 \beta ^2
   k^4-8 \beta ^2 k^4 m-6 \beta  k^2+6 \beta  k \omega
   -3 \gamma  l^2 -9 \alpha  \beta  B k^2 
   +\left(12 \beta ^2
   k^4 m-9 \alpha  A \beta  k^2 \right)\text{cn}^{2} \bigg]=0. 
\end{align}
Then, instead (\ref{con1}) we have 
\begin{equation} \label{con1B}
4 \beta ^2
   k^4-8 \beta ^2 k^4 m-6 \beta  k^2+6 \beta  k \omega
   -3 \gamma  l^2 -9 \alpha  \beta  B k^2 =0,
\end{equation}
whereas the second condition (\ref{con2}) remains the same.
From (\ref{con2}) and (\ref{con1B}) we get
\begin{equation} \label{A-omegaB}
A=\frac{4 k^{2} \beta}{3 \alpha} m, \qquad \omega = k+\frac{2 k^{3} \beta}{3}(2m-1)+\frac{l^{2}\gamma}{2 k \beta} +\frac{3 k \alpha}{2}B.
\end{equation}
Using (\ref{vcc1}) we have 
\begin{align} \label{bcn2}
 B & =- \frac{4\,k^{2}\beta}{3\alpha}\left(\frac{E(m)}{K(m)}+m-1\right)\qquad  \mbox{and} \\  \label{omegaB}
 \omega & = k-2 k^{3}\beta\left(\frac{E(m)}{K(m)}+\frac{m-2}{3}
 \right) +\frac{l^{2}\gamma}{2k \beta } .
\end{align}
Therefore the function $u(x,y,t)$ given by (\ref{cn2B}),
where $B$ is given by (\ref{vcc1}), is the  solution to (\ref{kdv21}) conserving the displaced fluid's volume (mass), for arbitrary $k,l$ and $m\in (0,1]$.


\subsection{Periodic superposition solutions}  \label{PSup}

Now, we check whether the superposition solution, first found by Khare and Saxena in \cite{KhSa13,KhSa14} for the KdV equation, can be the solution to  (2+1)-dimensional KdV equation (\ref{kdv21}).

\subsubsection{Mathematical solution}
Let us consider the surface profile function in the form
\begin{align} \label{supM}
u(x,y,t) & =\frac{A}{2} \bigg(\text{dn}^{2} [(k x+l y-\omega t),m] 
\pm\sqrt{m}\,\text{cn} [(k x+l y-\omega t),m]\,\text{dn} [(k x+l y-\omega t),m]\bigg),  
\end{align} 
Substituting (\ref{supM}) into (\ref{kdv21}) yields (after some simplifications)
\begin{equation} \label{supMex}
\frac{A\sqrt{m}}{24 k \beta}\, \left[ C_{0}+C_{\text{cn}2}\, \text{cn}^{2}+C_{\text{cn}\text{dn}}\, \text{cn}\, \text{dn}+C_{\text{cn}4}\, \text{cn}^{4}+C_{\text{cn}3\text{dn}}\, \text{cn}^{3}\, \text{dn}\right] = 0, 
\end{equation}
where the coefficients $C_{i}$ are the following
\begin{align} \label{c0}
C_{0} & = -\left((m-1) \left(3 \beta  k^2 (4-3 \alpha  A (m-1))+2 \beta ^2 k^4 (5
   m-1)-12 \beta  k \omega +6 \gamma  l^2\right)\right) \\ \label{c1}
C_{\text{cn}2} & = m \left(3 \beta  k^2 (8-15 \alpha  A (m-1))+8 \beta ^2 k^4 (7 m-5)-24 \beta k \omega +12 \gamma  l^2\right) \\ \label{c2}
C_{\text{cn}\text{dn}} & = \sqrt{m} \left(-3 \beta  k^2 (9 \alpha  A (m-1)-8)+16 \beta ^2 k^4 (2 m-1)-24 \beta  k \omega +12 \gamma  l^2\right) \\ \label{c3}
C_{\text{cn}4} & = 12 k^{2}m^{2}\beta(3A \alpha-4 k^{2}\beta) \quad \mbox{and} \quad 
C_{\text{cn}3\text{dn}} = 12 k^{2}m^{3/2}\beta(3A \alpha-4 k^{2}\beta).
\end{align} 
So, the equation (\ref{supMex}) is valid when the coefficients $C_{i}$ vanish simultaneously. Equations (\ref{c3}), $C_{\text{cn}4}\!=\!0$ and $C_{\text{cn}3\text{dn}}\!=\!0$, supply the same condition ~$\displaystyle A=\frac{4k^{2}\beta}{3\alpha}$. After insertion this form of $A$ into (\ref{c1}) and (\ref{c2}) we find that both of these equations are equivalent to
\begin{equation} \label{c1-2}
\beta ^2 k^4 (m-5)-6\beta  k^2+6\beta k\omega -3\gamma  l^2=0 \quad \Longrightarrow  \quad \omega=k+\frac{k^3 \beta}{6}(5-m)+\frac{l^{2}\gamma}{2 k \beta}.
\end{equation}
The equation  (\ref{c0})  with
\begin{equation} \label{ccc}
 A=\frac{4k^{2}\beta}{3\alpha} \qquad \mbox{and}\qquad \omega=k+\frac{k^3 \beta}{6}(5-m)+\frac{l^{2}\gamma}{2 k \beta}
\end{equation}
becomes the identity. So, the function (\ref{supM}) with $A,\omega$ given by (\ref{ccc}) is the {\bf mathematical} superposition solution to (2+1)-dimensional KdV equation (\ref{kdv21}), for arbitrary values of $k,l$ and $m\in(0,1]$.

\subsubsection{Physical solution}

In order to ensure volume conservation, take
\begin{align} \label{supF}
u(x,y,t) & =\frac{A}{2} \left(\text{dn}^{2} [(k x+l y-\omega t),m] 
\pm\sqrt{m}\,\text{cn} [(k x+l y-\omega t),m]\,\text{dn} [(k x+l y-\omega t),m]\right)+B. 
\end{align} 
Substituting (\ref{supF}) into (\ref{kdv21}) gives, after some simplifications, equation of the same form as (\ref{supMex}), but with slightly different coefficients $C_{i}$. Now, coefficients $C_{\text{cn}4}$ and $C_{\text{cn}3\text{dn}} $ stay the same as in (\ref{c2})-(\ref{c3}), but $C_{\text{0}}, C_{\text{cn}2}, C_{\text{cndn}}$ become $B$-dependent. 
Explicitly,
\begin{align} \label{c0F}
C_{0} & = (1-m) \left(3 \beta  k^2 (-3 \alpha  A (m-1)+6 \alpha  B+4)+2 \beta^2 k^4 (5 m-1)-12 \beta  k \omega +6 \gamma  l^2\right) \\ \label{c1F}
C_{\text{cn}2} & = m \left(3 \beta  k^2 (-15 \alpha  A (m-1)+12 \alpha  B+8)+8 \beta ^2 k^4 (7 m-5)-24 \beta  k \omega +12 \gamma  l^2\right)  \\ \label{c2F}
C_{\text{cn}\text{dn}} & = \sqrt{m} \left(3 \beta  k^2 (-9 \alpha  A (m-1)+12 \alpha  B+8)+16 \beta ^2 k^4 (2 m-1)-24 \beta  k \omega +12 \gamma  l^2\right).
\end{align}
Substitution $A=\frac{4k^{2}\beta}{3\alpha}$ into (\ref{c0F})-(\ref{c2F}) gives
\begin{align} \label{c0FA}
2 (m-1) \left(-3 \beta  k^2 (3 \alpha  B+2)+\beta ^2 k^4
   (m-5)+6 \beta  k \omega -3 \gamma  l^2\right) & = 0 \\ \label{c1FA}
-4 m \left(-3 \beta  k^2 (3 \alpha  B+2)+\beta ^2 k^4 (m-5)+6
   \beta  k \omega -3 \gamma  l^2\right)   & = 0 \\ \label{c2FA}
-4 \sqrt{m} \left(-3 \beta  k^2 (3 \alpha  B+2)+\beta ^2 k^4
   (m-5)+6 \beta  k \omega -3 \gamma  l^2\right)   & = 0. 
\end{align} 
For $m\in(0,1)$ we have only one condition 
\begin{align} \label{c02}
-3 \beta k^2 (3\alpha B+2)+\beta ^2 k^4 (m-5)+6 \beta k\omega -3\gamma l^2 & = 0.
\end{align} 

Volume conservation requires 
\begin{equation} \label{vcc} 
\int_{0}^{L_{x}} \int_{0}^{L_{y}}u(x,y,t)\, dx\,dy=0,
\end{equation}
Periodicity of the $(\text{dn}^{2}[(k x+l y-\omega t),m] \pm\sqrt{m}\,\text{cn} [(k x+l y-\omega t),m]\,\text{dn} [(k x+l y-\omega t),m])$ function implies 
$$ L_x=\frac{4 K(m)}{k} \qquad \mbox{and} \qquad  L_y=\frac{4 K(m)}{l} ,$$
where $K(m)$ is the complete elliptic integral of the first kind. So, from (\ref{vcc}) we obtain
\begin{align} \label{vcc1S}
B  =-\frac{1}{L_{x}L_{y}}  \int_{0}^{L_{x}}\!\! \int_{0}^{L_{y}} &  \frac{A}{2} \bigg(\text{dn}^{2} [(k x+l y-\omega t),m]  
\pm\sqrt{m}\,\text{cn} [(k x+l y-\omega t),m]\,\text{dn} [(k x+l y-\omega t),m]\bigg) \, dx\,dy \nonumber \\ &   \hspace{2ex} = -\frac{A\, E(m)}{2\,K(m)} =  -\frac{2\, k^{2}\beta}{3\alpha} \frac{E(m)}{K(m)}. 
\end{align} 
So finally, physical superposition solutions to (2+1)-dimensional KdV equation (\ref{kdv21}), have the form (\ref{supF}) with 
\begin{equation}
A=\frac{4k^{2}\beta}{3\alpha}, \quad B=-\frac{2\, k^{2}\beta}{3\alpha} \frac{E(m)}{K(m)}, \quad \omega=k -  k^3\beta \left(\frac{E(m)}{K(m)} +\frac{m-5}{6} \right)+\frac{\gamma l^2}{2\beta k}, \quad \mbox{for~arbitrary~} k,l \mbox{~and~}  m\in(0,1). 
\end{equation}

It is worth noting that due to properties of the Jacobi elliptic functions, the superpositions ~$u_{+}=\text{dn}^{2}+\sqrt{m}\,\text{cn}\,\text{dn}$~ and $u_{-}=\text{dn}^{2}-\sqrt{m}\,\text{cn}\,\text{dn}$~ represent the same wave, just shifted relative to each other by the period of the Jacobi function \cite{KhSa13,KhSa14}. 


\subsection{The Kadomtsev-Petviashvili type equation from ideal fluid model} \label{KPe}

For a flat bottom, (2+1)-dimensional KdV equation (\ref{kdv21}) implies the Kadomtsev-Petviashvili type equation. Differentiating (\ref{kdv21}) over $x$ yields
\begin{equation} \label{kpt}
\frac{\partial }{\partial x}\left(  \frac{\partial u}{\partial t} + \frac{\partial u}{\partial x} +\frac{3}{2} \alpha\, u \frac{\partial u}{\partial x} +\frac{1}{6} \beta\, \frac{\partial^{3} u}{\partial x^{3}}  \right)= -\lambda\, \frac{\partial^{2} u}{\partial y^{2}}, \quad \mbox{where} \quad \lambda=\frac{1}{2} \frac{\gamma}{\beta}. 
 \end{equation}
Equation (\ref{kpt}) can be named the {\bf Kadomtsev-Petviashvili equation in a fixed reference frame}. It is derived in first-order perturbation approach from the ideal fluid model, when $\alpha\approx\beta$ and $\gamma\approx\beta^{2}$.

The classical Kadomtsev-Petviashvili equation \cite{KP} has the following form
\begin{equation} \label{KPcl}
\frac{\partial }{\partial \hat{x}}\left(  \frac{\partial u}{\partial \hat{t}}   +6 u \frac{\partial u}{\partial \hat{x}} + \frac{\partial^{3} u}{\partial \hat{x}^{3}}  \right)= -\lambda\, \frac{\partial^{2} u}{\partial \hat{y}^{2}}.
\end{equation}
Equation (\ref{KPcl}) with $\lambda >0$ is called KP2 equation, whereas (\ref{KPcl}) with $\lambda <0$ is called KP1.

Let us make a scaling transformation of (\ref{kpt}) to a moving reference frame. Take 
\begin{equation} \label{scalxt}
\hat{x} =\sqrt{\frac{3}{2}}(x-t), \quad \hat{t} =\frac{1}{4} \sqrt{\frac{3}{2}}\,\alpha t, \quad \mbox{and} \quad \hat{y}= y.
\end{equation}
Then ~$\frac{\partial ~}{\partial x}=\sqrt{\frac{3}{2}}\frac{\partial ~}{\partial \hat{x}}$, ~$\frac{\partial u}{\partial t}=\sqrt{\frac{3}{2}}(\frac{\alpha}{4}\frac{\partial u}{\partial \hat{t}}-\frac{\partial u}{\partial \hat{x}})$, ~$\frac{\partial u}{\partial x}=\sqrt{\frac{3}{2}}\frac{\partial u}{\partial \hat{x}}$,  ~$\frac{\partial^{3} u}{\partial x^{3}}=\frac{3}{2}\sqrt{\frac{3}{2}}\frac{\partial^{3} u}{\partial \hat{x}^{3}}$~ and ~$\frac{\partial^{2} u}{\partial y^{2}}=\frac{\partial^{2} u}{\partial \hat{y}^{2}}$.\\

So, in the moving reference frame equation (\ref{kpt}) receives the form 
\begin{equation} \label{kpm}
\frac{\partial }{\partial \hat{x}}\left(  \frac{\partial u}{\partial \hat{t}}   +6 u \frac{\partial u}{\partial \hat{x}} +\frac{\beta}{\alpha} \frac{\partial^{3} u}{\partial \hat{x}^{3}}  \right)= -\lambda\, \frac{\partial^{2} u}{\partial \hat{y}^{2}}
 \end{equation}
where $\lambda =\frac{4}{3} \frac{\gamma}{\alpha\beta}$. 
In case $\alpha=\beta$, equation (\ref{kpm}) coincides with the classical  Kadomtsev-Petviashvili equation (\ref{KPcl}). To our best knowledge, this is the first derivation of the KP equation from the ideal fluid model.

The classical Kadomtsev-Petviashvili equation  (\ref{KPcl}) possesses several analytic solutions. It is well known 
that the function 
$u(\hat{x},\hat{y},\hat{t})=A\, \text{Sech}^{2} (k \hat{x}+l \hat{y}-\omega \hat{t})$
fulfils equation KP (\ref{KPcl}) when $A=2k^{2}$, $\omega= 4k^{3}+\lambda\, l^{2}/k$ and $k,l$ are arbitrary constants. Multi-soliton solutions and multi-lump solutions to KP equation are known \cite{YS2020,PeSt93,DT20,DT21,KM21,FCH22}, as well.
With a similar approach as that presented in subsection \ref{PSup}, we showed that superpositions of the form (\ref{supM}) or (\ref{supF}) are also the solutions to the KP equation (\ref{KPcl}) or (\ref{kpm}). For example the superposition (\ref{supF}) is the solution to KP equation (\ref{kpm}) for arbitrary $k,l$ when
$$A=\frac{2\beta\,k^{2}}{\alpha}, \quad B= -\frac{\beta\,k^{2}}{\alpha}\frac{E(m)}{K(m)}, \quad \omega=\frac{\beta\,k^{3}}{\alpha}\left(5-m-6  \frac{E(m)}{K(m)}\right).  $$

More details on this subject will be presented elsewhere.

\subsection{The one-dimensional motion of the obtained solutions} \label{1dimS}
All analytic solutions $u(x,y,t)$ considered in this Section depend on $x,y,t$ in a particular way, namely $u(x,y,t)=u(\xi)$, $\xi=kx+ly-\omega t$. Therefore, despite explicit dependence on $x,y$ coordinates, they describe 1-dimensional motion.  This property is easily seen when we introduce $x',y'$ coordinates, rotated to $x,y$ by the angle $\varphi=\arctan(\frac{l}{k})$. In these rotated coordinates we have
\begin{equation} \label{xyrot}
u(kx+ly-\omega t)=u\big(k(x'\cos\varphi-y'\sin\varphi)+l(x'\sin\varphi+y'\cos\varphi)-\omega t) \big)=u(\sqrt{k^{2}+l^{2}}\,x'-\omega t).
\end{equation}
The functions given by (\ref{xyrot}) clearly represent waves moving in the direction $x'$ with constant values in the perpendicular direction. 
In other words, the solutions of the form $u(x,y,t)=u(kx+ly-\omega t)$ exhibit translational symmetry in $y'$ direction.

\begin{figure} 
\resizebox{0.6\columnwidth}{!}{\includegraphics{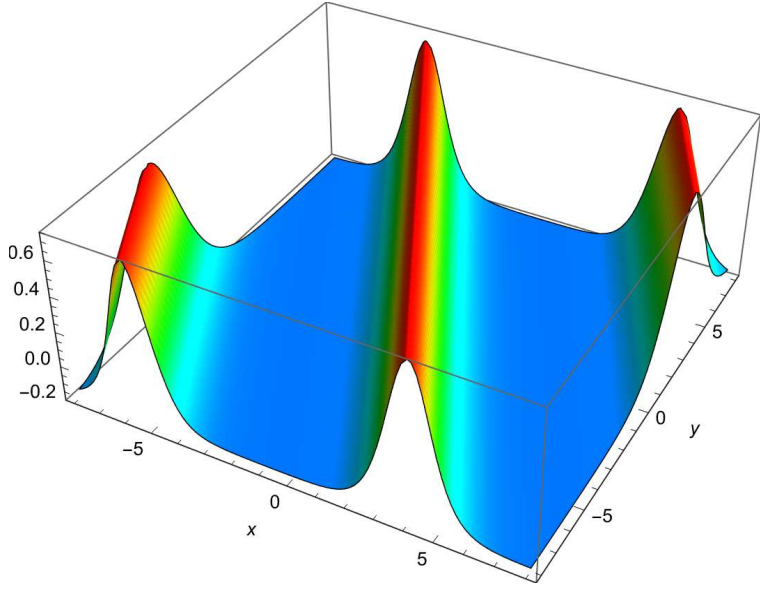}}
\caption{Snapshot of the cnoidal solution (\ref{cn2B}) for (2+1)-dimensional KdV equation in three dimensions. For parameters used in this example, see details in the text.  }\label{fff1}
\end{figure}

\begin{figure}[htb]
\begin{center}
\resizebox{0.6\columnwidth}{!}{\includegraphics{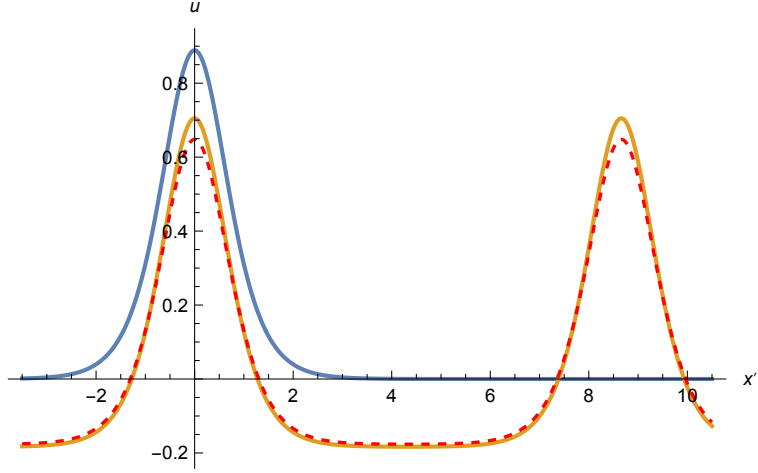}}
 \caption{Examples of solutions to (2+1)-dimensional KdV equation presented in rotated coordinates. Blue - soliton solution (\ref{ssol}), orange - cnoidal solution (\ref{cn2B}), red dashed - superposition solution (\ref{supF}). For parameters used, see details in the text.} \label{fff2}
\end{center}
 \end{figure}

Some examples of solutions obtained are illustrated in Figs.~\ref{fff1} and \ref{fff2}. 
In these examples we set $\alpha=0.15$, $\beta=0.1$, $\gamma=0.05$, $k=1$,  $l=0.5$. For the cnoidal solution (\ref{cn2B}), we took $m=0.999$, whereas, for the superposition solution, we took $m=0.85989$ to ensure the same wavelengths for both periodic solutions.

In Fig.~\ref{fff1} the snapshot of two-dimensional periodic cnoidal solution (\ref{cn2B}) is presented.

In Fig.~\ref{fff2} the wave profiles are presented in rotated (scaled) coordinates $x',y'$. 
In these scaled dimensionless variables the amplitudes of the presented three waves are as follows: $A_{\text{sol}}  \approx 0.88889,~A_{\text{cno}}\approx 0.88768,~A_{\text{sup}}\approx 0.82388$.
For cnoidal and superposition waves the amplitudes shown above are defined as $A_{(\cdot)}=u_{\text{crest}}-u_{\text{trough}}$. 
These waves propagate with the following dimensionless velocities: 
$v_{\text{sol}}  \approx 1.00996,~v_{\text{cno}}\approx 0.97299,~v_{\text{sup}}\approx 0.97007$.

Perhaps, more interesting are properties of these waves expressed in dimension variables. Assume shallow water case and take $H=10$ meters. 
Using inverse of the scaling transformation (\ref{scalxt}) we can find that wave amplitudes are $A_{\text{sol}} \approx 1.33333~\text{m},~~ A_{\text{cno}}\approx 1.33152~\text{m},~~\mbox{and}~~A_{\text{sup}}\approx 1.23581~\text{m} .$
The corresponding propagation speeds of these waves are as follows:
$ v_{\text{sol}} \approx 10.5443\, \frac{\text{m}}{\text{s}},~~v_{\text{cno}}\approx 10.1583\, \frac{\text{m}}{\text{s}},~~v_{\text{sup}}\approx 10.1278\, \frac{\text{m}}{\text{s}}$. 
The wavelengths of both periodic waves presented in Fig.~\ref{fff2} is 
$\lambda\approx 10.37$ m. 
\section{(2+1)-dimensional fifth-order KdV equation  \\ Case 6:  \hspace{1ex}  $\bal=O(\bbe^{2})$, \hspace{1ex} $\bga=O(\bbe^2)$, \hspace{1ex} $\bde=0$.  } \label{5thEq}

Denote now
\begin{equation} \label{ag}
\alpha= A\,\beta^{2}, \quad \gamma = G\,\beta^2, \end{equation}
where, as earlier, $A,G$ are arbitrary constants close to 1. 

In (1+1)-dimensional case, perturbation approach allowed to derive the fifth-order KdV equation \cite{BS2013}.

We consider the flat bottom case, $\delta=0$, aiming to derive second order wave equation. 
Therefore it is enough to use the velocity potential in the form 
\begin{align} \label{po8}
\phi = f & -\frac{1}{2} z^2 (\beta f_{2x}+\gamma f_{2y}) 
+ \frac{1}{24} z^4 (\beta^2 f_{4x}+2\beta\gamma f_{2x2y}+\gamma^2\,f_{4y}) 
\\ &  
 - \frac{1}{720} z^6 \left(\beta^3f_{6x}+3\beta^2\gamma f_{4x2y}+3\beta\gamma^{2}f_{2x4y}+\gamma^3 f_{6y}\right)
+ \cdots   \nonumber
\end{align}
neglecting all terms of higher order than $\beta^3$. Substituting velocity potential (\ref{po8}) into (\ref{G2}) and retaining only terms up to $\beta^2$
yields the first of Boussinesq's equations for this case as
\begin{align} \label{bu1-6}
u_{t} +f_{xx} &- \frac{1}{6}\beta\,f_{4x} +\frac{\gamma}{\beta}\, f_{yy} 
+\alpha \left(u f_{x}\right)_{x} +\frac{1}{120} \beta\, f_{6x} -\frac{1}{3}\gamma\,f_{2x2y}=0. 
\end{align}
Note that the last three terms in (\ref{bu1-6}) are second order ones.

The second Boussinesq's equation is obtained from (\ref{G3}). However, we will use this equation in the form which takes into account possible surface tension effects (see, \cite[Eqs.~(9)~and~(11)]{KRnody22})
\begin{align}  \label{G3st}
\phi_t + \frac{1}{2}\alpha \left(\phi_x^2+\frac{\gamma}{\beta}\phi_y^2+\frac{1}{\beta}\phi_z^2\right) + u{ -\tau (\beta u_{xx}+\gamma u_{yy})}&= 0,   
 \quad \mbox{for~} z =1+\alpha u ,
\end{align}
where $\tau=\frac{T}{\varrho g H^{2}}$ is the Bond number ($T$ is surface tension coefficient, $\varrho$ is fluid density, $g$ is gravitational acceleration and $H$ is fluid depth).

So, we substitute the velocity potential (\ref{po8}) into (\ref{G3st}) and retain only terms up to $\beta^2$. This gives
$$ u + f_{t} -\beta\left( \frac{1}{2}\,f_{2xt}+\tau u_{2x} \right) +\frac{1}{2} \alpha f_{x}^{2} +\frac{1}{24}\beta^{2} f_{4xt} -\gamma\left(\frac{1}{2} f_{2yt}+\tau u_{2y}\right) =0. $$
After differentiation over $x$ the resulting second  Boussinesq's equation receives the form 
\begin{equation} \label{bu2-6}
u_{x}+f_{xt} -\beta\left( \frac{1}{2}\,f_{3xt}+\tau u_{3x} \right) +\alpha\, f_{x} f_{2x}+\frac{1}{24}\beta^{2}\,f_{5xt}-\gamma\left(\frac{1}{2}f_{x2yt}+\tau u_{x2y}\right) =0.
\end{equation}  
Because Boussinesq's equations (\ref{bu1-6}) and (\ref{bu2-6}) do not contain $y$ dependence in zeroth order terms, we can follow the standard method used in section \ref{stand-g2}. Denoting $f_{x}=w$, $f=\int w dx$, and so on, we can write equations (\ref{bu1-6}) and (\ref{bu2-6}) as
\begin{align} \label{b1-6}
u_{t} +w_{x} & - \frac{1}{6}\beta\,w_{3x} +\frac{\gamma}{\beta} \int w_{yy}\,dx
+\alpha \left(u w\right)_{x} +\frac{1}{120} \beta^{2}\, w_{5x} -\frac{1}{3}\gamma\,w_{x2y}=0, \\ \label{b2-6}
w_{t} +u_{x} & -\beta\left(\frac{1}{2}\,w_{2xt}+\tau u_{3x} \right) +\alpha\, w w_{x}+\frac{1}{24}\beta^{2}\,w_{4xt}-\gamma\left(\frac{1}{2} w_{2yt} +\tau u_{x2y}\right)=0.
\end{align} 
Note that the last three terms in both equations are second order ones.

Zeroth-order equations (\ref{b1-6})-(\ref{b2-6}) are the same as those of (\ref{et})-(\ref{wt}). Therefore in zeroth-order relations (\ref{0g2}) hold. In first order we have terms with $\beta$ and $\frac{\gamma}{\beta}$, so we seek for $w$ in the form
\begin{equation} \label{1wa}
w =u+w^{(1)} = u +\beta Q^{(b)}+\frac{\gamma}{\beta} Q^{(gb)}.
\end{equation}
Proceeding in the same way as in section \ref{stand-g2} we obtain correction  functions in first order as
\begin{equation} \label{p1-5}
Q^{(b)}=\left(\frac{2-3\tau}{6} \right) u_{xx}\,, \qquad Q^{(gb)}=
 - \frac{1}{2}\int\left(\int u_{yy}\,dx\right) dx.
\end{equation}
So, with $w= u +\beta \left(\frac{2-3\tau}{6} \right) u_{xx}-  \frac{1}{2} \frac{\gamma}{\beta} \int\left(\int \textcolor{blue}{u_{yy}} \,dx\right) dx$ we obtain first order 
compatibility of Boussinesq's equations (\ref{b1-6})-(\ref{b2-6}) as
\begin{equation} \label{p1-5a}
u_{t} +u_{x} +\beta \left(\frac{1-3\tau}{6}\right)u_{3x}+\frac{\gamma}{2 \beta} \int u_{yy}\,dx =0 .
\end{equation}
Now, we consider second order corrections. We seek for $w$ in the form
\begin{align} \label{2wa}
w =u+w^{(1)}+w^{(2)} = u & +\beta  \left(\frac{2-3\tau}{6} \right) u_{xx}-  \frac{1}{2} \frac{\gamma}{\beta} \int\left(\int u_{yy}\,dx\right) dx \\ & +\frac{1}{4}
\alpha\,u^{2} + \beta^{2}\frac{12-20\tau-15\tau^{2}}{120} u_{4x}+\gamma \,Q^{(g)} +\frac{\gamma^{2}}{\beta^{2}} Q^{(bg)}. \nonumber
\end{align}
Second order correction terms proportional to $\alpha$ and $\beta^{2}$, that is $\frac{1}{4}\alpha\,u^{2} + \beta^{2}\frac{12-20\tau-15\tau^{2}}{120} u_{4x}$ are already known from (1+1)-dimensional case (see, \cite{BS2013}).
Substituting $w$ given by (\ref{2wa}) into (\ref{b1-6}), and retaining terms up to second order we obtain
\begin{align} \label{2wa1}
u_{t} +u_{x}& +\beta \left(\frac{1-3\tau}{6}\right)u_{3x}+\frac{\gamma}{2 \beta}\int u_{yy}\,dx +\frac{3}{2}\alpha u u_{x}  
+\beta^{2} \left(\frac{19-30\tau-45\tau^{2}}{360}\right)u_{5x} \\ &
+\gamma\left(Q^{(g)}_{x} +\left(\frac{1-6\tau}{12}\right) u_{x2y} \right)  +\frac{\gamma^{2}}{\beta^{2}} \left(Q^{(bg)}_{x} -
\frac{1}{2}\int (\int(\int u_{4y}\,dx)\,dx)\,dx \right)=0.  \nonumber 
\end{align} 
The same steps with equation (\ref{b2-6}) give
\begin{align} \label{2wa2}
u_{t} +u_{x}& -\beta \left(\frac{1+3\tau}{6}u_{2xt}+\tau u_{3x}\right)-\frac{\gamma}{2 \beta}\int ( \int u_{yyt}\,dx)\,dx +\alpha\left(-\frac{1}{2}u u_{t} +u u_{x}\right)  \\ & +\beta^{2}\left(\frac{-3+10\tau-15\tau^{2}}{120}\,u_{4xt}\right)+\gamma\left(Q^{(g)}_{t}-\frac{1}{4} u_{yyt}-\tau u_{xyy}
 \right) +  \frac{\gamma^{2}}{\beta^{2}}\left(Q^{(bg)}_{t} \right) =0.  \nonumber
\end{align}
Now, we have to replace $t$-derivatives by $x$-derivatives. From (\ref{p1-5}),
$u_{t} =-u_{x} +\beta \left(\frac{1-3\tau}{6}\right)u_{3x}+\frac{\gamma}{2 \beta} \int u_{yy}\,dx$, valid up to first order. This expression serves for replacements in terms $u_{xxt}, u_{yyt}$ which appear in first order terms of (\ref{2wa2}). In second order terms of (\ref{2wa2}) it is enough to use zeroth order approximation, that is $u_{t}=-u_{x}, u_{4xt}=-u_{5x}$,  $u_{2yt}=-u_{x2y}$, and $Q^{(g)}_{t}=-Q^{(g)}_{x}, Q^{(bg)}_{t}=-Q^{(bg)}_{x}$.  The result, valid up to second order is
\begin{align} \label{2wa3}
u_{t} +u_{x}& +\beta \left(\frac{1-3\tau}{6}\right)u_{3x} +\frac{\gamma}{2 \beta} \int u_{yy}\,dx +\frac{3}{2}\alpha u u_{x} +\beta^{2} \left(\frac{19-30\tau-45\tau^{2}}{360}\right)u_{5x} \\ & +\gamma
\left(-Q^{(g)}_{x}+\left(\frac{5}{12}-\tau\right) u_{xyy}\right) +  \frac{\gamma^{2}}{\beta^{2}}\left(-Q^{(bg)}_{x}+\frac{1}{4}\int (\int (\int u_{4y}\,dx)\,dx)\,dx \right) =0.  \nonumber
\end{align}
Equations (\ref{2wa1}) and (\ref{2wa3}) should be compatible. Substracting (\ref{2wa3}) from (\ref{2wa1}) yields
\begin{align} \label{2wa4}
\gamma\left(2 Q^{(g)}_{x} -\left(\frac{1}{3}+ \frac{1}{2}\tau\right)u_{xyy} \right) + \frac{\gamma^{2}}{\beta^{2}}\left(2 Q^{(bg)}_{x}- \frac{3}{4}\int (\int (\int u_{4y}\,dx)\,dx)\,dx  \right)&=0,
\end{align}
what, due to the independence of small parameters is equivalent to two separate equations. Integrating them one obtains yet unknown corrections as
\begin{equation}
Q^{(g)}=\frac{2-3\tau}{12} u_{yy}\qquad \mbox{and} \qquad Q^{(bg)}=  
\frac{3}{8}\int (\int (\int (\int u_{4y}\,dx)\,dx)\,dx)\,dx  .
\end{equation}
In this way, with the following $w$
\begin{align} \label{2waFin}
w = u & +\beta  \left(\frac{2-3\tau}{6} \right) u_{xx}-\frac{\gamma}{\beta}  \frac{1}{2} \int\left(\int u_{yy}\,dx\right) dx  
+ \alpha\frac{1}{4}\,u^{2} + \beta^{2}\left(\frac{12-20\tau-15\tau^{2}}{120}\right) u_{4x} \\ & +\gamma \left(\frac{2-3\tau}{12}\right) u_{yy}  +\frac{\gamma^{2}}{\beta^{2}}\, \frac{3}{8}\int (\int (\int (\int u_{4y}\,dx)\,dx)\,dx)\,dx \nonumber
\end{align}
we obtain compatibility of the Boussinesq equations (\ref{b1-6})-(\ref{b2-6}).
Both receive the same form 
\begin{align} \label{2wa12F}
u_{t} +u_{x}& +\beta \left(\frac{1-3\tau}{6}\right)u_{3x}+\frac{\gamma}{2\beta} \!\int u_{yy}\,dx + \frac{3}{2}\alpha\, u u_{x}  
+\beta^{2} \left(\frac{19-30\tau-45\tau^{2}}{360}\right)u_{5x} \\ &
+\gamma\left(\frac{1-3\tau}{4} \right) u_{x2y} -
\frac{1}{8} \frac{\gamma^{2}}{\beta^{2}} \!\int (\int(\int u_{4y}\,dx)\,dx)\,dx =0,  \nonumber 
\end{align} 
{\bf The equation (\ref{2wa12F}) is indeed the (2+1)-dimensional extension of the fifth-order Korteweg-de Vries equation derived from the Euler equations describing the irrotational motion of an ideal fluid.}

The equation (\ref{2wa12F}) is nonlocal. By differentiating it over $x$ we can obtain an analogue to  Kadomtsev-Petviashvili equation
\begin{equation} \label{kp5}
\frac{\partial }{\partial x} \left( \frac{\partial u}{\partial t} + \frac{\partial u}{\partial x}  +\beta\frac{(1\!-\!3\tau)}{6} \, \frac{\partial^{3} u}{\partial x^{3}} +\frac{3}{2} \alpha\, u \frac{\partial u}{\partial x}
+\frac{19\!-\!30\tau\!-\!15\tau^{2}}{360}\beta^{2}\frac{\partial^{5} u}{\partial x^{5}} \right)= -\frac{\gamma}{2\beta}\, \frac{\partial^{2} u}{\partial y^{2}} \textcolor{blue}{~-\frac{\gamma^{2}}{8\beta^{2}} \int(\int u_{4y}\,dx)\,dx}.
 \end{equation}
 According to our best knowledge, this is the first equation of Kadomtsev-Petviashvili - type related to the (2+1)-dimensional fifth-order KdV equation. Note that the last term (blue) is the second order term. If this term can be neglected (or if $u_{4y}=0$) then the analogy to the KP equation (\ref{kpt}) or (\ref{kpm}) is complete. For shallow water case, when we can neglect surface tension effects (because $\tau\in (10^{-8},10^{-6}$) this equations  simplifies to
 $$ \frac{\partial }{\partial x} \left( \frac{\partial u}{\partial t} + \frac{\partial u}{\partial x}  +\frac{1}{6}\beta \, \frac{\partial^{3} u}{\partial x^{3}} +\frac{3}{2} \alpha\, u \frac{\partial u}{\partial x}
+\beta^{2}\frac{\partial^{5} u}{\partial x^{5}} \right)= -\frac{\gamma}{2\beta}\, \frac{\partial^{2} u}{\partial y^{2}}.
 $$

\section{(2+1)-dimensional Gardner equation \\ Case 7:  \hspace{1ex}  $\bbe=O(\bal^{2})$, \hspace{1ex} $\bga=O(\bal^3)$, \hspace{1ex} $\bde=0$.  } \label{GardEq}

Denote now 
\begin{equation} \label{ag2a}
\beta= B\,\alpha^{2}, \quad \gamma = G\,\alpha^3, \end{equation}
where, as earlier, $B,G$ are arbitrary constants close to 1. We consider flat bottom case, $\delta=0$, aiming to derive second order wave equation. In (1+1)-dimensional case perturbation approach allowed to derive the Gardner  equation \cite{BS2013}.

The velocity potential (\ref{pot8}) can be rewritten (up to fourth order) as
\begin{align} \label{pot8G3}
\phi = f & -\frac{1}{2} z^2  (\alpha^{2}B f_{2x}+ \alpha^{3}G f_{2y}) 
+ \frac{1}{24} z^{4}\alpha^{4} B^{2} f_{4x}+ \cdots
\end{align}
Substituting velocity potential (\ref{pot8G3}) into (\ref{G2}) and retaining terms up to second order yields 
\begin{equation} \label{BGar13}
u_{t} + f_{xx} +\frac{\gamma}{\beta} f_{yy} +\alpha (u f_{x})_{x} + \frac{\alpha\gamma}{\beta} (u f_{y})_{y} -\frac{1}{6} \beta f_{4x}
=0.
\end{equation}
Note, that term with $\frac{\gamma}{\beta}$ is now first order one, and terms 
with $\frac{\alpha\gamma}{\beta}$ and $\beta$ are of second order.

In the same way, from (\ref{G3}) we obtain 
\begin{equation} \label{BGar23}
u +f_{t} + \frac{1}{2}\alpha f_{x}^{2} +\frac{1}{2}\frac{\alpha\gamma}{\beta}  f_{y}^{2} -\frac{1}{2}\beta\left(f_{xxt} +2\tau u_{xx}\right) 
=0.
\end{equation}

Equations (\ref{BGar13}) and (\ref{BGar23}) are (2+1)-dimensional (second order) Boussinesq's equations for the case $\beta\approx\alpha^{2}, \gamma\approx\alpha^{3}$. In zeroth order they coincide with those which lead to KdV, so we can follow the same method that was used in sections \ref{stand-g2} and \ref{5thEq}.

 Denoting $f_{x}=w$, $f=\int w dx$, and so on, and differentiating (\ref{BGar23}) we can write equations (\ref{BGar13})-(\ref{BGar23}) as
\begin{align} \label{BGa13}
u_{t} + w_{x} & +\frac{\gamma}{\beta}\!\int\! w_{yy}\,dx +\alpha (u w)_{x} + \frac{\alpha\gamma}{\beta}\left(u_{y} \!\int\! w_{y}\,dx +u\!\int\! w_{yy}\,dx \right)-\frac{1}{6} \beta w_{3x} =0,\\  \label{BGa23}
w_{t} + u_{x} & + \alpha w w_{x} +\frac{\alpha\gamma}{\beta}\, w_{y}\! \int \!w_{y}dx - \frac{1}{2}\beta\left( w_{xxt} +2\tau u_{xxx}\right) =0.
\end{align}

Zeroth-order equations (\ref{BGa13})-(\ref{BGa23}) are the same as those of (\ref{et})-(\ref{wt}). Therefore in zeroth-order relations (\ref{0g2}) hold. In first order we have terms with $\alpha$ and $\frac{\gamma}{\beta}$, so we seek for $w$ in the form
\begin{equation} \label{1waGa}
w =u+w^{(1)} = u +\alpha Q^{(a)}+\frac{\gamma}{\beta} Q^{(gb)}.
\end{equation}

Proceeding in the same way as in section \ref{stand-g2} we obtain correction  functions in first order as
\begin{equation} \label{pGa}
Q^{(a)}=-\frac{1}{4}u^{2}\,, \qquad Q^{(gb)}=
 - \frac{1}{2}\int\left(\int u_{yy}\,dx\right) dx.
\end{equation}
So, in first order perturbation approach we obtain the wave equation
\begin{equation} \label{G3rz1}
u_{t} + u_{x} +\frac{3}{2}\alpha\,u u_{x}+ \frac{1}{2}\frac{\gamma}{\beta}\!\int\! u_{yy}dx =0.
\end{equation}
Seeking for second order wave equation we take (correction function $\frac{\alpha^{2}}{8}u^{3}$ is known from (1+1)-dimensional theory, see, \cite{BS2013})
\begin{equation} \label{2waGa}
w =u+w^{(1)} +w^{(2)} = u - \frac{\alpha}{4} u^{2} +\frac{\beta}{6}(2-3\tau)u_{xx} -\frac{1}{2}\frac{\gamma}{\beta}\int\left(\int u_{yy}\,dx\right)dx 
+\frac{\alpha^{2}}{8} u^{3}+ \frac{\alpha\gamma}{\beta}Q^{(agb)}+\frac{\beta^{2}}{\alpha^{2}} Q^{(ba2)}+\frac{\gamma^{2}}{\beta^{2}} Q^{(gb2)}.
\end{equation}
Then from (\ref{BGa13}) we obtain (retaining terms up to second order)
\begin{align} \label{BGa13q}
u_{t} + u_{x} & +\frac{3}{2}\alpha uu_{x} +\frac{\gamma}{2\beta}\!\int\! u_{yy}dx -\frac{3}{8} \alpha^{2} u^{2}u_{x}
+\frac{1}{6}\beta\left(1-3\tau\right) u_{3x}  \\ &  +
 \frac{\alpha\gamma}{\beta}\left(Q^{(agb)}_{x} -\frac{1}{2}\!\int\! (u u_{y})_{y}\,dx +\frac{1}{2} u\!\int\! u_{yy}\,dx +3u\!\int\!u^{2}u_{yy}\,dx  +u_{y}\!\int\! u_{y}\,dx  - \frac{1}{2}u_{x}\!\int\!(\!\int\! u_{yy} dx)\,dx\right)  \nonumber\\ & + \frac{\beta^{2}}{\alpha^{2}}Q^{(ba2)}_{x}
 + \frac{\gamma^{2}}{\beta^{2}}\left(Q^{(gb2)}_{x} -\frac{1}{2}\int (\int (\int u_{4y}\,dx)\,dx)\,dx\right) =0, \nonumber 
\end{align}
whereas from (\ref{BGa23}) the result is 
\begin{align} \label{BGa23q}
u_{t} + u_{x} & +\alpha \left(u u_{x}-\frac{1}{2} u u_{t}\right) 
-\frac{\gamma}{2\beta}\!\int\! ( \!\int\! u_{yyt}\,dx)\,dx +\alpha^{2}\left(-\frac{3}{4} u^{2}u_{x}+\frac{3}{8} u^{2}u_{t} \right) -\beta\left(\frac{1-3\tau}{6} u_{xxt} +\tau u_{3x}\right) \\ &+
 \frac{\alpha\gamma}{\beta}\left(Q^{(agb)}_{t} -\frac{1}{2}u\!\int\! u_{yy}\,dx\! +u_{y}\!\int\! u_{y}\,dx -\frac{1}{2}u_{x}\!\int\!(\!\int\! u_{yy}\,dx)\,dx\right)  + \frac{\beta^{2}}{\alpha^{2}} Q^{(ba2)}_{t}
+ \frac{\gamma^{2}}{\beta^{2}} Q^{(gb2)}_{t} =0.\nonumber 
\end{align}

Now, we have to replace $t$-derivatives by $x$-derivatives. From (\ref{G3rz1}),
$u_{t} =-\left(u_{x}+\frac{3}{2}\alpha u u_{x} +\frac{\gamma}{2 \beta} \int u_{yy}\,dx\right)$, valid up to first order. This expression serves for replacements in terms $u_{t}, u_{xxt}$ which appear in first order terms of (\ref{BGa23q}). In second order terms of (\ref{BGa23q}) it is enough to use zeroth order approximation, that is $u_{t}=-u_{x}, u_{xxt}=-u_{3x}$,  $u_{2yt}=-u_{x2y}$, and $Q^{(agb)}_{t}=-Q^{(agb)}_{x}, Q^{(ba2)}_{t}=-Q^{(ba2)}_{x}$, $Q^{(gb2)}_{t}=-Q^{(gb2)}_{x}$.  The result, valid up to second order is
\begin{align} \label{BG102}
u_{t} + u_{x} & +\frac{3}{2}\alpha uu_{x} +\frac{\gamma}{2\beta}\!\int\! u_{yy}dx -\frac{3}{8} \alpha^{2} u^{2}u_{x}
+\frac{1}{6}\beta\left(1-3\tau\right) u_{3x} \\ & +
 \frac{\alpha\gamma}{\beta}\left(-Q^{(agb)}_{x} +\frac{3}{4}\!\int\! (u u_{y})_{y}\,dx -\frac{1}{4} u \!\int\! u_{yy}\,dx + u_{ y}\!\int\! u_{y}\,dx -u_{x}\!\int\!(\!\int\! u_{yy}\,dx)\,dx \right) \nonumber \\ & +
 \frac{\beta^{2}}{\alpha^{2}} \left(-Q^{(ba2)}_{x}\right) +  \frac{\gamma^{2}}{\beta^{2}} \left(-Q^{(gb2)}_{x} + \frac{1}{4} \int (\int (\int u_{4y}\,dx)\,dx)\,dx \right) =0.
\end{align}
Substraction of (\ref{BG102}) from (\ref{BGa13q}) yields
\begin{align} \label{BGdiff}
 & \frac{\alpha\gamma}{\beta}\left(2Q^{(agb)}_{x} - \frac{5}{4}\!\int\! (u_{y}^{2}+u u_{yy})\,dx\! + \frac{3}{4}u\!\int\! u_{yy}\,dx +3 u \!\int\! u^{2} u_{yy}\,dx\right)  + \frac{\beta^{2}}{\alpha^{2}} \left(2 Q^{(ba2)}_{x}\right)
\\ & + \frac{\gamma^{2}}{\beta^{2}} \left(2Q^{(gb2)}_{x} - \frac{3}{4} \int (\int (\int u_{4y}\,dx)\,dx)\,dx \right)
=0. \nonumber
\end{align}
This equation give three independent conditions on $Q^{(agb)}_{x}, Q^{(ba2)}_{x}$, and $Q^{(gb2)}_{x}$. Integrating them one obtains correction functions as 
\begin{align} \label{BGcor}
Q^{(agb)} & = \frac{5}{8}\!\int\!( \!\int\! (u_{y}^{2}+u u_{yy})\,dx)dx - \frac{3}{8}\!\int\! (u\!\int\! u_{yy}\,dx)\,dx - \frac{3}{2}\!\int\! ( u \!\int\! u^{2} u_{yy} \,dx)\,dx \nonumber \\ 
Q^{(ba2)} & = 0, \qquad\qquad Q^{(gb2)} = \frac{3}{8} \int (\int (\int (\int u_{4y}\,dx)\,dx)\,dx)\,dx .
\end{align}

With these correction functions Boussinesq's equations (\ref{BGa13})-(\ref{BGa23}) become compatible. The final wave equation  receives the following form
\begin{align} \label{BGarEq}
u_{t} + u_{x} & +\frac{3}{2}\alpha uu_{x} +\frac{\gamma}{2\beta}\!\int\! u_{yy}dx - \frac{3}{8}\alpha^{2} u^{2}u_{x}
+\frac{1}{6} \beta (1-3\tau) u_{3x} \nonumber  \\ &+
 \frac{\alpha\gamma}{\beta}\left(\frac{1}{8} \!\int\! (u_{y}^2 +u u_{yy})\,dx+\frac{1}{8}u \!\int\! u_{yy}\,dx +\frac{3}{2} u \!\int\! u^2 u_{yy}\,dx+u_{y}\!\int\! u_{y}\,dx -\frac{1}{2} u_{x}
   \!\int\! (\!\int\! u_{yy} \, dx)\,dx \right) \\ &
- \frac{1}{8} \frac{\gamma^{2}}{\beta^{2}}  \!\int\! (\!\int\! (\!\int\! u_{4y}\,dx)\,dx)\,dx =0.  \nonumber 
\end{align}
Note that when $u=u(x,t)$ does not depend on $y$, equation (\ref{BGarEq}) reduces to
\begin{equation} \label{GARD}
u_{t} + u_{x} +\frac{3}{2}\alpha uu_{x} - \frac{3}{8}\alpha^{2} u^{2}u_{x}
+\beta \frac{1-3\tau}{6} u_{3x} =0,
\end{equation}
which is well known Gardner equation.
Therefore, equation (\ref{GARD}), the simplest extension of the Gardner equation to (2+1)-dimensions can be called {\bf (2+1)-dimensional Gardner equation}.

The partial integro-differential equation (\ref{BGarEq}) is highly nonlocal. Even differentiating over $x$ several times does not yield a local partial differential equation.

\section{Extensions to cases of uneven bottom} \label{bottom}

In \cite{KRcnsns}, we showed in (1+1)-dimensional theory that small variations of the bottom introduce additional terms in Boussinesq's equations. These terms can be either first or second order depending on the relation of $\delta$ parameter to the leading parameter $\beta$ (or $\alpha$ in the case leading to the Gardner equation). However, in each case, the Boussinesq equations could be reduced to single wave equations for surface profile function $u(x,t)$, and the additional term in such equations have the universal form. It turned out that in each of the KdV, fifth-order KdV and Gardner equations extension to the case of uneven bottom adds the term $-\frac{1}{4}\delta (2h u_{x}+h_{x} u)$. However, compatibility is possible only if $h_{xx}=0$ (for details, see \cite{KRcnsns}).

It turns out that in (2+1)-dimensional theory all these results hold. In section \ref{stand-g2} we derived equation (\ref{et11}) which is (2+1)-dimensional KdV equation taking into account small bottom variations.
Analogous procedures (not presented here) including $\delta\ne 0$ for the case
discussed in section \ref{5thEq} lead to the equation
\begin{align} \label{2wa1D}
u_{t} +u_{x}& +\beta \left(\frac{1-3\tau}{6}\right)u_{3x}+\frac{\gamma}{2\beta} \!\int u_{yy}\,dx + \frac{3}{2}\alpha\, u u_{x}  
+\beta^{2} \left(\frac{19-30\tau-45\tau^{2}}{360}\right)u_{5x} \\ &
+\gamma\left(\frac{1-3\tau}{4} \right) u_{x2y} -
\frac{1}{8} \frac{\gamma^{2}}{\beta^{2}} \!\int (\int(\int u_{4y}\,dx)\,dx)\,dx -\frac{1}{4}\delta (2h u_{x}+h_{x} u)  =0,  \nonumber 
\end{align} 
which differs from (\ref{2wa12F}) only by the last term ~$-\frac{1}{4}\delta (2h u_{x}+h_{x} u)$. Note, that all terms in the second row are of second order. Equation (\ref{2wa1D}) is (2+1)-dimensional 5th order KdV equation taking into account small bottom variations. For more details, see the Appendix A.

When the case $\delta\approx\alpha^{2}$ is considered together with $\beta\approx\alpha^{2},~ \gamma\approx\alpha^{3}$ (like in section \ref{GardEq}) the resulting equation is
\begin{align} \label{BGarEqD}
u_{t} + u_{x} & +\frac{3}{2}\alpha uu_{x} +\frac{\gamma}{2\beta}\!\int\! u_{yy}dx - \frac{3}{8}\alpha^{2} u^{2}u_{x}
+\frac{1}{6} \beta (1-3\tau) u_{3x} \nonumber \\ &+
 \frac{\alpha\gamma}{\beta}\left(\frac{1}{8} \!\int\! (u_{y}^2 +u u_{yy})\,dx+\frac{1}{8}u \!\int\! u_{yy}\,dx +\frac{3}{2} u \!\int\! u^2 u_{yy}\,dx+u_{y}\!\int\! u_{y}\,dx -\frac{1}{2} u_{x}
   \!\int\! (\!\int\! u_{yy} \, dx)\,dx \right) \\ &
- \frac{1}{8} \frac{\gamma^{2}}{\beta^{2}}  \!\int\! (\!\int\! (\!\int\! u_{4y}\,dx)\,dx)\,dx -\frac{1}{4}\delta (2h u_{x}+h_{x} u) =0.  \nonumber 
\end{align}
Equation (\ref{BGarEqD}) is (2+1)-dimensional Gardner equation taking into account small bottom variations. It differs from the case with the flat bottom (\ref{BGarEq}) by the universal term ~$-\frac{1}{4}\delta (2h u_{x}+h_{x} u)$. For more details, see the Appendix~B.

\section{Conclusions}\label{concl}

We have shown that for some particular relations between small parameters, it is possible to rigorously derive (2+1)-dimensional nonlinear wave equations of KdV-type from the ideal fluid model. 
The crucial condition allowing for such derivations is that the parameter 
$\gamma$, responsible for wavelength in the direction perpendicular to the main wave propagation, has to be much smaller than $\beta$ parameter related to the wavelength of the main wave. In such a case, zeroth-order Boussinesq's equations do not contain $y$-derivatives. This fact make it possible to use relations (\ref{0g2}) between $x$- and $t$-derivatives of the wave profile function that are true in zeroth order. Then, when $\alpha\approx\beta, \gamma\approx\beta^{2}$, limiting to first-order perturbation approach, we obtain (2+1)-dimensional nonlocal extension to KdV equation. After transformation to a  moving reference frame and differentiation over a new $x$-variable, this equation becomes the Kadomtsev-Petviashvili equation.  
It appears that the derived (2+1)-dimensional nonlocal KdV equation possesses for sure several families of analytic solutions: soliton, cnoidal and superposition solutions which are essentially (1+1)-dimensional. 

We derived (2+1)-dimensional nonlocal extension to the fifth-order KdV equation when $\alpha\approx\gamma\approx\beta^{2}$  
and (2+1)-dimensional nonlocal extension to the Gardner equation when $\beta\approx\alpha^{2}, \gamma\approx\alpha^{3}$. All these equations are obtained without any additional assumptions.

In all cases discussed in the present paper, the effects of the uneven bottom can be included if the bottom function fulfills the condition $h_{xx}=0$. The related term, which considers small variations of the bottom, has a universal form. 

\appendix
\section{5th-order KdV equation with an uneven bottom}
Denote now 
\begin{equation} \label{ag2}
\alpha= A\,\beta^{2}, \quad \gamma = G\,\beta^2, \quad \delta= D\,\beta^2. \end{equation}
In this case, the boundary condition at the bottom (\ref{G4}) imposes the following relation
\begin{align} \label{r9c5}
F & = \beta^{3} D \left(h_x f_x +h f_{xx}\right) + \beta^{4}DG \left(h_y f_y +h f_{yy} \right)+\beta^{5} D^{2}\left(h h_{x} F_{x} -\frac{1}{2}h^{2}F_{xx}\right)+O(\beta^{6}) . \end{align}
Therefore we can eliminate $F$ using 
\begin{align} \label{r9c5a}
F & = \beta^{3} D \left(h_x f_x +h f_{xx}\right) + \beta^{4}DG \left(h_y f_y +h f_{yy} \right),  \end{align}
valid up to fourth order. Since we aim to derive wave equation valid up to second order in small parameters it is enough to keep terms in the velocity potential up to third order. This gives
\begin{align} \label{po8-5}
\phi = f & -\frac{1}{2} z^2 (\beta  f_{2x}+ \beta^{2}\, G  f_{2y}) 
+ \frac{1}{24} z^4 (\beta^2  f_{4x}+2\beta^{3}\, G f_{2x2y})  \\ &
 - \frac{1}{720} z^6 \beta^3 \,f_{6x} 
 + z  \beta^{3} D \left(h_x f_x +h f_{xx}\right) + O(\beta^{4}). \nonumber
\end{align}
Substituting the velocity potential (\ref{po8-5}) into the kinematic boundary condition at the surface (\ref{G2}) and retaining terms up to second order yields equation (\ref{bu1-6}) supplemented by term ~$-\delta (h f_{x})_{x}$.
Like in other cases (see, e.g., equation (\ref{wt}) and all cases studied in \cite{KRcnsns}) no $\delta$ dependent term appear in the Boussinesq equation originating  from 
the dynamic boundary condition at the surface (\ref{G3}). 
Then the obtained bottom dependent term in the wave equation (\ref{2wa12F}) has the same form ~$-\frac{1}{4}\delta(2h u_{x}+h_{x} u)$.

\section{Gardner equation with an uneven bottom}
Now we have
\begin{equation} \label{ag3a}
\beta= B\,\alpha^{2}, \quad \gamma = G\,\alpha^3, \quad \delta= D\,\alpha^2. 
\end{equation}
The boundary condition at the bottom (\ref{G4}) implies
\begin{align} \label{r9cG}
F & = \alpha^{4} D \left(h_x f_x +h f_{xx}\right) + \alpha^{5}DG \left(h_y f_y +h f_{yy} \right)+O(\alpha^{6}) , \end{align}
which allows us to express $F$ by $f$ and its derivatives. Then the velocity potential taken up to fourth order in $\alpha$ is
\begin{align} \label{po8-G}
\phi = f & -\frac{1}{2} z^2 \alpha^{2} (B f_{2x}+ \alpha\, G  f_{2y}) 
+ \frac{1}{24} z^4 \alpha^4  B^{2} f_{4x} 
 + z  \alpha^{4}B D \left(h f_x\right)_{x} + O(\beta^{5}) .
\end{align}
In this case, to obtain Boussinesq's equations valid up to second order, velocity potential valid to fourth order is necessary (due to term $\frac{1}{\beta}\approx\frac{1}{\alpha^{2}}$ in (\ref{G3})). Then with the velocity potential (\ref{po8-G}), from (\ref{G2}) we obtain equation (\ref{BGar13}) supplemented by the term $-\delta (h f_{x})_{x}$. From (\ref{G3}) the result is equation (\ref{BGar23}) without any change. Next, with the same procedure we arrive to equation (\ref{BGarEqD}).


\end{document}